\documentclass[%
 reprint,
 showpacs,preprintnumbers,
 amsmath,amssymb,
 aps,
]{revtex4-1}

\usepackage{graphicx}
\usepackage{dcolumn}
\usepackage{bm}
\usepackage{braket}
\usepackage[version=3]{mhchem}

\newcommand{\figref}[1]{Fig.~\ref{#1}}

\newcommand{\secref}[1]{Sec.~\ref{#1}}

\newcommand{\mrm}{\mathrm}

\newcommand{\mcl}{\mathcal}

\newcommand{\br}[1]{\left( #1 \right)}

\newcommand{\BR}[1]{\left[ #1 \right]}

\newcommand{\hspc}{\hspace{1em}}

\newcommand{\uar}{\uparrow}
\newcommand{\dar}{\downarrow}

\newcommand{\im}{\mrm{i}}
\newcommand{\ex}{\mrm{e}}
\newcommand{\alp}{\alpha}
\newcommand{\bet}{\beta}
\newcommand{\gam}{\gamma}
\newcommand{\del}{\delta}
\newcommand{\eps}{\epsilon}

\newcommand{\lam}{\lambda}
\newcommand{\sig}{\sigma}

\newcommand{\ome}{\omega}

\newcommand{\Del}{\Delta}

\newcommand{\ham}{\mcl{H}}
\newcommand{\vam}{\mcl{V}}
\newcommand{\pjc}{\mcl{P}}
\newcommand{\qjc}{\mcl{Q}}
\newcommand{\pri}{\prime}
\newcommand{\dpri}{{\prime\prime}}
\newcommand{\Mn}{m}

\newcommand{\rsat}{\mrm{sat}}
\usepackage{color,ulem}

\begin{document}
\title{Magnetic excitations in magnetization plateaux of a frustrated spin ladder}
\author{Kota Sasaki}
\email{1517617@alumni.tus.ac.jp}
\affiliation{Department of Applied Physics, Tokyo University of Science, Katsushika, Tokyo 125-8585, Japan}
\author{Takanori Sugimoto}
\email{sugimoto.takanori@rs.tus.ac.jp}
\affiliation{Department of Applied Physics, Tokyo University of Science, Katsushika, Tokyo 125-8585, Japan}
\author{Shigetoshi Sota}
\affiliation{Computational Materials Science Research Team, RIKEN Center for Computational Science, Kobe, Hyogo 650-0047, Japan}
\author{Takami Tohyama}
\affiliation{Department of Applied Physics, Tokyo University of Science, Katsushika, Tokyo 125-8585, Japan}

\date{\today}

\begin{abstract}
  Magnetization plateaux emerging in quantum spin systems due to spontaneously breaking of translational symmetry have been reported both theoretically and experimentally.
  The broken symmetry can induce reconstruction of elementary excitations such as Goldstone and Higgs modes, whereas its microscopic mechanism and reconstructed quasi-particle in magnetization-plateau phases have remained unclear so far.
  Here we theoretically study magnetic excitations in the magnetization-plateau phases of a frustrated spin ladder by using dynamical density-matrix renormalization-group method. Additionally, analytical approaches with perturbation theory are performed to obtain intuitive view of magnetic excitations.
 Comparison between numerical and analytical results indicates the presence of a reconstructed quasi-particle originating from spontaneously broken translational symmetry, which is realized as a collective mode of spin trimer called trimeron.

\end{abstract}

\pacs{Valid PACS appear here}
\maketitle

\section{INTRODUCTION}\label{intro}

Various quantum spin systems with frustration have been extensively studied, motivated by exotic characters such as quantum spin liquid at zero temperature and quantization of magnetization with spontaneously broken translational symmetry. Actually, gapless quantum spin liquid states and gapped quantized states of magnetization have been reported not only theoretically but experimentally~\cite{balents10,oshikawa97,totsuka98,kageyama99}. These states are often induced by frustration, and are switchable by applied magnetic field. For example, zigzag spin chain, where geometrical frustration originates from antiferromagnetic first- and second-neighbor interactions, is known as a typical quantum spin system exhibiting a gapped-to-gapless transition induced by magnetic field at zero temperature~\cite{okunishi03}.

As compared with the ground-state properties, dynamical behaviors in magnetic fields almost have not been clarified so far. In particular, dynamical properties in the quantized state of magnetization, so-called magnetization plateau (MP) state, mostly remains unclear, despite possible emergence of novel elementary excitations due to spontaneous symmetry breaking. In fact, recent studies on a weakly-coupled spin-ladder compound have reported a Higgs mode due to spontaneously broken symmetries~\cite{hong17,hong17-2,ying19}. Furthermore, these dynamical behaviors are crucial for understanding spin/heat transport, which is applicable to spintronics devices~\cite{uchida08,hirobe16}.

In this paper, we focus on magnetic excitations in a frustrated spin ladder (FSL), where antiferromagnetic interactions are assigned to the first- and second-neighbor bonds in a leg and the first-neighbor bond in a rung. This model exhibits three MPs at normalized finite magnetization $\Mn\equiv M/M_\rsat= 1/3$, $1/2$, and $2/3$ with saturation magnetization $M_\rsat$~\cite{sugimoto15,sugimoto18}. Interestingly, all of these MPs are induced by spontaneously breaking of translational symmetry, so that the MPs exhibit extended magnetic unit cells that are different from the original unit cell of this model. In addition, this model is regarded as an effective spin model to reproduce magnetic behaviors in real materials \ce{BiCu2PO6}~\cite{abraham94,koteswararao07,mentre09,tsirlin10} and \ce{Li2Cu2O(SO4)2}~\cite{rousse17,sun15,vaccarelli17,vaccarelli19}. Actually in \ce{BiCu2PO6}, external field dependences~\cite{kohama12,casola13,kohama14,colmont18}, dynamical properties~\cite{plumb13,plumb16}, and thermal conductivity~\cite{nagasawa14,jeon16,kawamata18} observed experimentally have been theoretically explained in the FSL model~\cite{lavarelo11,sugimoto13,shyiko13}, though additional terms such as Dzyaloshinskii-Moriya interaction are required to obtain a quantitative coincidence~\cite{splinter16,hwang16}. Therefore, the FSL model deserves to be investigated in terms of the relation between low-energy excitations and spontaneously broken symmetries of MP phases.

The preceding studies on the MP states~\cite{sugimoto15,sugimoto18} have presented equivalence of two different models, the FSL in the strong rung limit and an anisotropic frustrated spin chain (AFSC). According to these studies, the $\Mn=1/3$, $1/2$, and $2/3$ MP states in the FSL correspond to $\Mn^\pri=-1/3$, $0$, and $1/3$ MP states in the AFSC, respectively. Since the $\Mn^\pri=-1/3$ and $\Mn^\pri=1/3$ MP states in the AFSC are connected with each other via a spin-flip pair, the corresponding $\Mn=1/3$ MP and $\Mn=2/3$ states in the FSL should have a common origin. Therefore, the dynamics of the $\Mn=1/3$ MP state is expected to be equivalent to that of the $\Mn=2/3$ MP state, while the $\Mn=1/2$ MP state can show qualitatively-different dynamics.

We perform numerical calculations of dynamical spin structure factor (DSSF) by using the dynamical density-matrix renormalization group (DDMRG) method~\cite{white92,jeckelmann02,schollwock05,sota10} to clarify the difference of dynamics in the $\Mn=1/3$, $1/2$, and $2/3$ MP states for the FSL in the strong-rung limit.
The dynamical behaviors of the $\Mn^\pri=0$ and $1/3$ MP states in the AFSC are also examined as compared with the dynamics of the FSL. The AFSC is useful for intuitive understanding of dynamical properties because of its simplicity.
Moreover, a perturbative clusterization approach imposing spontaneously breaking of translational symmetry is used to obtain intuitive physical pictures of spin dynamics.

The contents of this paper are as follows.
In Sec.~II, we introduce the model Hamiltonians of the FSL and the AFSC. The equivalence of the two different models is briefly reviewed with a projection operator to low-lying states in the strong rung limit of the FSL. We also introduce the DSSFs and model parameters for calculation.
In Sec.~III, the DSSFs obtained by the DDMRG are shown for three MP states in the FSL and two MP states in the AFSC. Section IV is used to give qualitative explanation of characteristic structures in the DSSFs and intuitive physical picture of spin dynamics.
For this sake, we introduce a perturbative clusterization approach imposing spontaneously breaking of translational symmetry. Finally, we summarize our results in Sec.~V.

\section{MODEL AND METHOD}\label{modelmethod}

In this section, we introduce two model Hamiltonians: the FSL and its corresponding model in the strong rung limit, the AFSC.
Additionally, the DSSFs that we calculate to investigate dynamical properties are defined.

\subsection{Frustrated spin ladder (FSL)}
The Hamiltonian of the FSL is defined as
\begin{align}
  \ham=\ham_{\perp}+\ham_{\parallel}+\ham_Z,
\label{H}
\end{align}
with
\begin{align}
    \ham_{\perp}&=J_{\perp}\sum_{i=1}^{N}{\bm S}_{i,1}\cdot{\bm S}_{i,2} \\
    \ham_{\parallel}&=\sum_{\eta=1,2}J_{\eta}\sum_i\sum_{j=1,2}{\bm S}_{i,j}\cdot{\bm S}_{i+\eta,j} \label{Hp}\\
    \ham_{Z}&=-H\sum_i\sum_{j=1,2}S^z_{i,j}
\end{align}
where ${\bm S}_{i,1}~({\bm S}_{i,2})$ is the $S=1/2$ spin operator on $i$th rung in the upper (lower) chain.
Exchange energies of the first-neighbor bond in a leg, the second-neighbor bond in a leg, and the first-neighbor bond in a rung, are denoted by $J_1$, $J_2$, and $J_\perp$, respectively. The magnitude of magnetic field is represented by $H$.
In this paper, we focus on the strong rung region of the FSL, because three MPs at $\Mn=1/3$, $1/2$, and $2/3$ become robust in this limit.
Moreover, this limit enables us to map the FSL to the AFSC, which is used to obtain an intuitive picture of dynamical behaviors.

\subsection{The effective model of an FSL: AFSC}
The Hamiltonian of AFSC is given by the bond-operator (quasi-spin) transformation~\cite{sugimoto15,sugimoto18,sachdev90,giamarchi99}.
To obtain the AFSC Hamiltonian with quasi-spin operators, we use the basis of singlet and triplet states on $i$th rung:
\begin{align}
  \ket{s}_i =& \frac{1}{\sqrt{2}}\br{\ket{\uar}_{i,1}\ket{\dar}_{i,2} - \ket{\dar}_{i,1}\ket{\uar}_{i,2}} \\
  \ket{t^+}_i=&\ket{\uar}_{i,1}\ket{\uar}_{i,2} \\
  \ket{t^0}_i=& \frac{1}{\sqrt{2}}\br{\ket{\uar}_{i,1}\ket{\dar}_{i,2} + \ket{\dar}_{i,1}\ket{\uar}_{i,2}} \\
  \ket{t^-}_i=&\ket{\dar}_{i,1}\ket{\dar}_{i,2}.
\end{align}
For simplicity, we call $\ket{t^\alpha}_i$ ($\alpha = +,0,-$) state ``$\alpha$ triplet'' in the following.
The Hamiltonian $\ham$ (\ref{H}) is rewritten by these bases.

In the strong rung limit, leg interactions in $\ham_\parallel$ (\ref{Hp}) are regarded as perturbative terms.
For $\ham_\parallel=0$, magnetization $M$ jumps from zero to saturation magnetization $M_\rsat$ at critical magnetic field $H_c=J_\perp$.
Finite contributions from $\ham_\parallel$ change the magnetization jump into a continuous curve including plateaux around the critical field. The range of field, $\Del H$, for partially-magnetized states approximately equals to $\Del H\sim J_\parallel$.
In this field region, the $0$ and $-$ triplets are much higher in energy than the $+$ triplet.
Therefore, we can ignore the $0$ and $-$ triplets, and thus obtain a low-energy effective Hamiltonian.
To abandon the high-energy triplets, we introduce the projection operator $\pjc=\prod_i\br{\ket{s}_i\bra{s}_i+\ket{t^+}_i\bra{t^+}_i}$.
The effective Hamiltonian is given by
\begin{align}
  \ham^\pri =& \mcl{P}\ham\mcl{P} \notag \\
  =& \sum_{\eta=1,2}\sum_{i} \frac{J_\eta}{2}(T_{i}^{+}T_{i+\eta}^{-}+T_{i}^{-}T_{i+\eta}^{+}+T_{i}^{z}T_{i+\eta}^{z})\notag \\
  &-H^\pri\sum_i T_i^z + \mrm{const.} \label{hamp},
\end{align}
where spin-$1/2$ quasi-spin operators at site $i$ are denoted by ${\bm T}_i$ given by $T^+_i=\ket{t^+}_i\bra{s}_i$, $T^-_i=\ket{s}_i\bra{t^+}_i$, and $T^z_i=\ket{t^+}_i\bra{t^+}_i-1/2$. The effective magnetic field $H^\pri$ is defined by $H^\pri=H-J_\perp-(J_1+J_2)/2$.
The effective Hamiltonian \eqref{hamp} describes AFSC.
Note that the $z$ component of quasi spin is given by $T^z_i=\ket{t^+}_i\bra{t^+}_i-1/2$. This leads to the relation that a normalized magnetization $\Mn^\pri\equiv M^\pri/M^\pri_\rsat=2\Mn-1$, $M^\pri$ ($M^\pri_\rsat$) being magnetization (saturation magnetization) in the quasi-spin system. For example, $\Mn=1/3$, $1/2$, and $2/3$ in the FSL correspond to $\Mn^\pri=-1/3$, $0$, and $1/3$ in the AFSC, respectively.

\subsection{Dynamical spin structure factor (DSSF)}
To investigate magnetic excitations of the FSL, we calculate the DSSF defined by
\begin{align}
  S^\pm({\bm q},\ome) = -\frac{1}{\pi}\Im\bra{\psi_0}S_{{\bm q}}^\mp\frac{1}{\omega-\ham+E_0+\im\gam}S_{{\bm q}}^\pm\ket{\psi_0}, \label{DSSF}
\end{align}
where $\ket{\psi_0}$ is the ground state, $E_0$ is the ground-state energy, and $\gam$ is an infinitesimal value.
The Fourier component $S^\pm_{\bm q}$ under the open boundary condition is given by
\begin{align}
  S^\pm_{\bm q} = \sqrt{\frac{2}{N+1}} \sum_i \sin(q_x i)S_{i,q_y}^\pm \label{Sq}
\end{align}
with
\begin{align}
  S^\pm_{i,q_y=0} = \frac{1}{\sqrt{2}}\br{S_{i,1}^\pm+S_{i,2}^\pm}, \hspc S^\pm_{i,q_y=\pi} = \frac{1}{\sqrt{2}}\br{S_{i,1}^\pm-S_{i,2}^\pm}, \label{Si}
\end{align}
where $S_{i,j}^\pm=S_{i,j}^x\pm i S_{i,j}^y$ and the wave number of the leg (rung) direction is given by $q_x=\frac{\pi}{N+1}n$ ($q_y=0,\pi$) with $n=1,2,\cdots,N$, $N$ being the total number of rung along the leg direction.

Similarly, the DSSF for AFSC denoted by $T^\pm(q,\ome)$ is given by substituting the quasi-spin operator $T_{q}^\pm$ and $\ham^\pri$ for $S_{{\bm q}}^\pm$ and $\ham$, respectively, in Eq.~\eqref{DSSF}. Here, $T_{q}^\pm=\sqrt{\frac{2}{N+1}} \sum_i \sin(q i)T_i^\pm$ and $q=\frac{\pi}{N+1}n$ with $n=1,2,\cdots,N$, $N$ being the total number of site in the chain.

To obtain the DSSF numerically, we use the DDMRG method~\cite{jeckelmann02,schollwock05}.
This method requires three target states, $\ket{\psi_0}$, $S^{\pm}_{\bf q}\ket{\psi_0}$ and $[\omega-\ham+E_0+\im\gamma]^{-1}S^{\pm}_{\bf q}\ket{\psi_0}$.
The correction vector $[\omega-\ham+E_0+\im\gamma]^{-1}S^{\pm}_{\bf q}\ket{\psi_0}$ is obtained by kernel-polynomial expansion method~\cite{sota10}.
In this method, a Gaussian broadening with a width $\sig$ is introduced instead of Lorentzian broadening in Eq.~(\ref{DSSF}).

\section{RESULT}\label{result}
$S^\pm_{\bm q}$ in Eq.~\eqref{Sq} has two modes, $q_y=0$ and $q_y=\pi$, with respect to rung parity.
These two modes in Eq.~\eqref{Si} are rewritten by using the singlet and triplet bases;
\begin{align}
  S^{\pm}_{i,q_y=0} &= \ket{t^\pm}_i\bra{t^0}_i + \ket{t^0}_i\bra{t^\mp}_i, \\
  S^{\pm}_{i,q_y=\pi} &= -\br{\ket{t^\pm}_i\bra{s}_i + \ket{s}_i\bra{t^\mp}_i}.
\end{align}
Since the $0$ triplet $\ket{t^0}_i$ in plateau regions is much higher in energy than the singlet $\ket{s}_i$ in the strong rung limit, only the $q_y=\pi$ mode is enough to describe elementary excitations in the low-energy region.
We thus discuss only the $q_y=\pi$ mode in this paper, and abbreviate $S^\pm(q_x,q_y=\pi,\omega)$ for FSL to $S^\pm(q_x,\omega)$ hereafter.

In our calculations, we use the following parameters: $J_1/J_\perp=0.2$ and $J_2/J_1=0.65$ in the 48-rung FSL and $\sig=0.02J_\perp$.
These parameters are chosen for the following reasons. Firstly, in the real material \ce{BiCu2PO6}, preceding studies have concluded that two spins on each rung form a singlet at low temperatures, i.e., the ground state without magnetic fields is so-called the rung-singlet state~\cite{plumb13,plumb16,lavarelo11,sugimoto13}.
The strong rung condition $J_1/J_\perp=0.2$ also belongs to the rung-singlet phase, and we can apply perturbation analysis based on the strong-rung limit to clarify an intuitive physical picture of elementary excitations.
Additionally, to stabilize the $m=1/3$, $1/2$, and $2/3$ MPs, we introduce the frustration $J_2/J_1=0.65$.
In fact, we have confirmed that the MPs emerge due to spontaneously breaking symmetry [see \figref{curve}(a) in Appendix A].
The system size, $N=48$ rungs, is sufficient to discuss dynamical behaviors at least qualitatively.
With these parameters, the truncation error in DDMRG is less than $10^{-4}$ with 600 states kept in the DDMRG calculations.

\subsection{$S^+$ excitation in the $\Mn=\dfrac{1}{2}$ MP phase}

\begin{figure*}[htbp]
 \begin{center}
  \includegraphics[height=55mm]{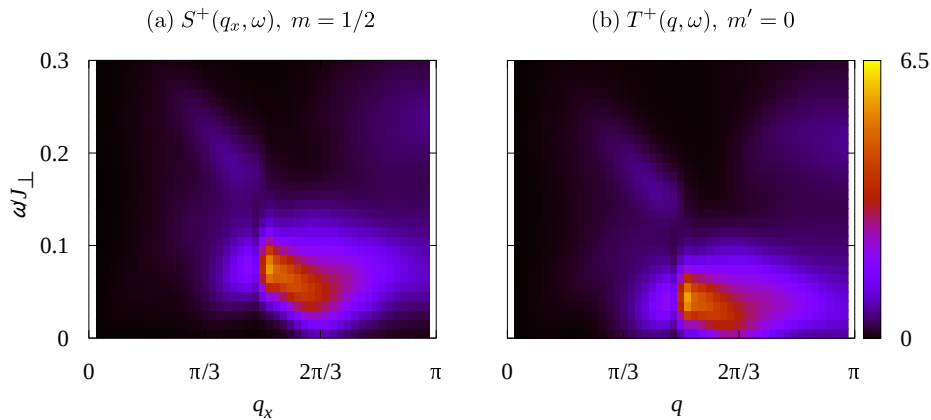}
 \end{center}
 \caption{
 (a) $S^+(q_x,\omega)$ in the $\Mn=1/2$ MP phase of FSL.
 (b) $T^+(q_x,\omega)$ in the $\Mn^\pri=0$ MP phase of AFSC.
A broad excitation with minimum energy excitations around $q_x=2\pi/3$ and an intensive peak at $q_x=\pi/2$ are common in (a) and (b).
 }
 \label{24}
\end{figure*}

Figure 1(a) shows $S^+(q_x,\omega)$ in the $\Mn=1/2$ MP phase [for MP see in \figref{curve}(a) in Appendix A].
A broad but intensive peak centered at $\omega/J_\perp=0.08$ is seen at $q_x=\pi/2$.
Its peak width is wider than the Gaussian width $\sig$, indicating intrinsic broadening of the peak.
We consider that this peak originates from a dimerized ground state in the $\Mn=1/2$ MP phase, which breaks the translational symmetry spontaneously~\cite{sugimoto15} and causes a doubled period of lattice.
A broad but dispersive structure extends above $q_x=\pi/2$ with minimum-energy excitations around $q_x=2\pi/3$.
This structure indicates multi-spinon excitation, and thus we consider it as a manifestation of fractionalized excitation with strong quantum fluctuation.
In fact, the ground state of the $\Mn=1/2$ MP in the FSL corresponds to the dimer state of $\Mn^\pri=0$ MP in the AFSC and thus we can interpret the broad excitation as multi spinons in the AFSC~\cite{lavarelo14}.
To confirm this interpretation, we calculate the DSSF of AFSC $T^+(q,\omega)$ in the $\Mn^\pri=0$ MP phase [see \figref{24}(b)].
As compared with \figref{24}(a), we find similar structures in \figref{24}(b): a broad excitation with minimum energy excitations around $q_x=2\pi/3$ and an intensive peak at $q_x=\pi/2$.
Therefore, we conclude that low-energy excitations are due to multi quasi-spinon excitations including the intensive peak caused by the doubled period of lattice.

\begin{figure*}[t]
 \begin{center}
  \includegraphics[height=55mm]{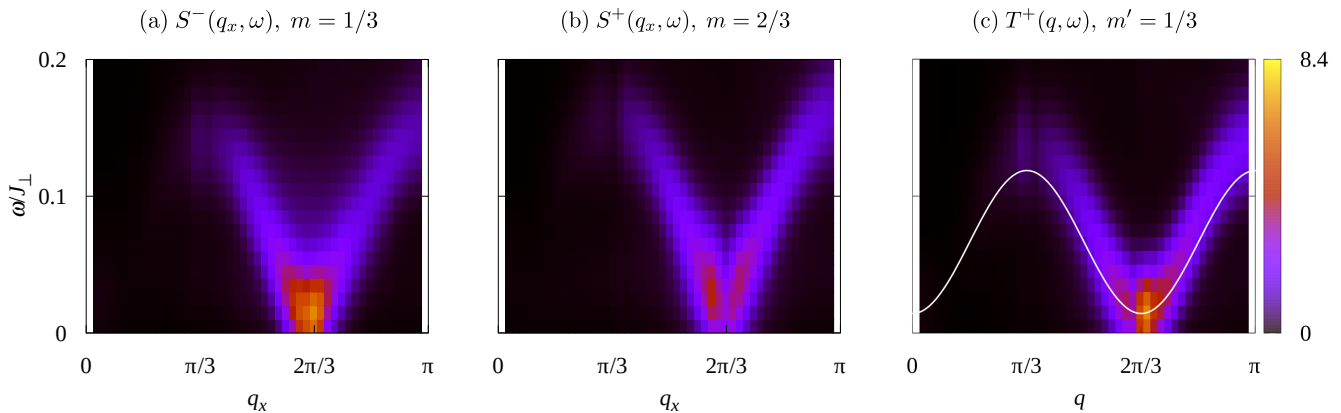}
 \end{center}
 \caption{
  (a) $S^-(q_x,\omega)$ in the $\Mn=1/3$ MP phase of FSL.
  (b) $S^+(q_x,\omega)$ in the $\Mn=2/3$ MP phase of FSL.
  (c) $T^+(q_x,\omega)$ in the $\Mn^\pri=1/3$ MP phase of AFSC, .
  The white line shows the dispersion relation obtained by an analytical calculation assuming clusterization (See \secref{discussion}).
 }
 \label{32}
\end{figure*}

\subsection{$S^-\, (S^+)$ excitation in the $\Mn=\dfrac{1}{3}\, \br{\dfrac{2}{3}}$ MP phase}\label{dssf2}
The $S^-(q_x,\omega)$ in the $\Mn=1/3$ MP phase and $S^+(q_x,\omega)$ in the $\Mn=2/3$ MP phase are shown in Figs.~\ref{32}(a) and \ref{32}(b), respectively.
These two spectra show a similar behavior with a dispersive feature with zero-energy excitation at $q_x=2\pi/3$, indicating a period with three times of the original unit-cell length in the real space.
This similarity is actually expected from the fact that both the $\Mn=1/3$ and $\Mn=2/3$ MPs can be associated with the array of quasi-spinons and share a common origin each other~\cite{sugimoto15}.
Since the $\Mn=2/3$ MP corresponds to the $\Mn^\pri=1/3$ MP,  the DSSF of ASFC, $T^+(q,\omega)$, in the $\Mn^\pri=1/3$ MP phase also shows a similar dispersive feature with zero-energy excitation at $q_x=2\pi/3$ as shown in \figref{32}(c).
Based on the similarity, we may construct an intuitive view of spin dynamics via full examination of $T^+(q,\omega)$ in the $\Mn^\pri=1/3$ MP phase.
We will discuss this view in \secref{discussion} using a clusterization approach.

\subsection{$S^+\, (S^-)$ excitation in the $\Mn=\dfrac{1}{3}\, \br{\dfrac{2}{3}}$ MP phase}
\begin{figure*}[t]
 \begin{center}
  \includegraphics[height=58mm]{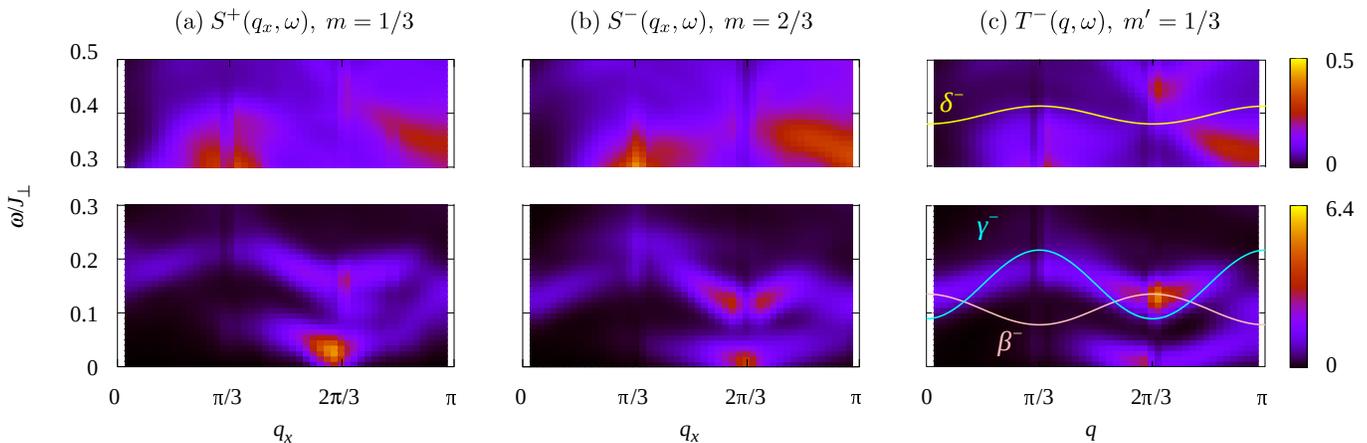}
 \end{center}
 \caption{
  (a) $S^+(q_x,\omega)$ in the $\Mn=1/3$ MP phase of FSL.
  (b) $S^-(q_x,\omega)$ in the $\Mn=1/3$ MP phase of FSL.
  (c) $T^-(q_x,\omega)$ in the $\Mn^\pri=1/3$ MP phase of AFSC.
  Every figure is split into upper and lower panels to make the distribution of spectrum visible.
  In (c), the pink line shows the dispersion of $\beta^-$ excitation obtained by an analytical calculation assuming clusterization
(See \secref{discussion}).
  The blue and yellow lines show the dispersion relations of $\gamma^-$ and $\delta^-$, respectively.
 }
 \label{16}
\end{figure*}

Figures~\ref{16}(a) and \ref{16}(b) show $S^+(q_x,\omega)$ in the $\Mn=1/3$ MP phase and $S^-(q_x,\omega)$ in the $\Mn=2/3$ MP phase, respectively.
These two figures share common features: two dispersive low-energy ($\omega/J_\perp<0.3$)  excitations with the minimum-energy excitation at $q_x=2\pi/3$ and high-energy broad excitations at $\omega/J_\perp>0.3$.
We find that $T^-(q,\omega)$ in the $\Mn^\pri=1/3$ MP phase shown in \figref{16}(c) exhibits spectral distributions similar to Figs.~\ref{16}(a) and \ref{16}(b).
To understand the origin of the spectra, we will introduce a clusterization approach for $T^-(q,\omega)$ in \secref{discussion}.

\section{DISCUSSION} \label{discussion}
Our purpose in this section is to give an intuitive physical view of elementary excitations in the $\Mn=1/3$ and $2/3$ MP phases using an analytical approach.
The following discussion is based on spontaneous translational symmetry breaking in the MP phases, where the magnetic unit cell is larger than the original unit cell.
In such a case, quantum entanglement between the magnetic unit cells is expected to be suppressed because of the enlargement of unit cell.
Therefore, effective interactions between the enlarged unit cells can be approximated to a semi-classical one [see Appendix B].
If the interactions are totally classical, the ground state is given by the direct product of local quantum states, which are obtained by exact diagonalization of local Hamiltonian in the enlarged unit cell.
Such a local quantum state contributing to the ground state should be one of low-lying states in the local Hamiltonian.
Otherwise, the inter-cell interactions will become larger than intra-cell interaction, contradicting with localized nature of spins in the magnetic unit cell.
Even in the case of semi-classical interactions where the ground state becomes a superposition of the direct product of local states, the local states contributing to the ground state should be low-lying states.
Based on this reasoning, we restrict the Hilbert space of the enlarged unit cell to several low-lying states obtained by exact diagonalization of local Hamiltonian.
Moreover, the interactions between the enlarged unit cells are projected onto the restricted Hilbert space.
We call this approach a clusterization based on spontaneously broken symmetry (CBSBS).

\begin{figure}[t]
 \begin{center}
  \includegraphics[width=70mm]{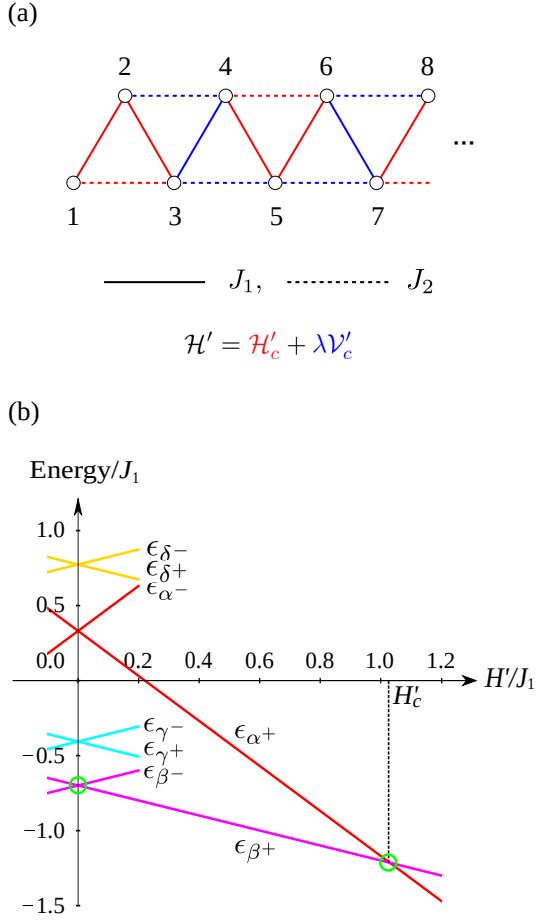}
 \end{center}
 \caption{
 (a) Clusterization based on spontaneously broken symmetry (CBSBS) of AFSC. The numbers denote the quasi-spin site. The solid (dotted) lines correspond to $J_1$ $(J_2)$ interactions. Red and blue colors of the lines are used to distinguish intra-cluster Hamiltonian $\ham_c^\pri$ and inter-cluster Hamiltonian $\vam_c^\pri$, respectively.
 (b) Energy levels of local states $\ket{\chi^\pm}$ in effective magnetic fields $H^\pri$ in the case of $J_2/J_1=0.65$. The lowest-energy state encounters level cross at  $H^\pri=0$ and $H_c^\pri$ denoted by the green circles.
 }
 \label{3spin}
\end{figure}

In the following, we apply the CBSBS to the $\Mn^\pri=1/3$ MP phase in the AFSC, because magnetic excitations in the $\Mn=1/3$ and $2/3$ MP phases of FSL are qualitatively similar to the $T^\pm$ excitations in the $\Mn^\pri=1/3$ MP phase of AFSC (see~\secref{result}).
The magnetic unit cell is enlarged three times as long as the original one due to spontaneous translational symmetry breaking. Therefore, we use the following Hamiltonian instead of Eq.~\eqref{hamp}:
\begin{equation}
  \ham^\pri = \ham_c^\pri + \lam\vam_c^\pri
\label{Heff}
\end{equation}
with
\begin{align}
  \ham_c^\pri&=\frac{J_1}{2}\sum_{i=1,2 (\mrm{mod}\,3)} D_1(i)+\frac{J_2}{2}\sum_{i=1 (\mrm{mod}\,3)} D_2(i) -H^\pri\sum_j T_j^z, \label{intra} \\
  \vam_c^\pri&=\frac{J_1}{2}\sum_{i=0 (\mrm{mod}\,3)} D_1(i)+\frac{J_2}{2}\sum_{i=0,2 (\mrm{mod}\,3)} D_2(i), \label{inter}
\end{align}
where the $\eta$th neighbor two-spin (dimer) interaction is represented by $D_\eta(i)=T_i^+T_{i+\eta}^- + T_i^-T_{i+\eta}^+ + T^z_iT_{i+\eta}^z$.
$\ham_c^\pri$ ($\vam_c^\pri$) corresponds to the intra (inter) cluster Hamiltonian denoted by the red (blue) lines in \figref{3spin}(a).
We introduce the coupling strength $\lam$ to control the inter-cluster interactions.
We note that Eq.~\eqref{Heff} with $\lam=1$ is equivalent to Eq.~\eqref{hamp}.

\subsection{Eigenstates of magnetic unit cell}
Since the Hamiltonian \eqref{intra} does not include the interaction between the clusters, we can diagonalize it in each cluster.
The resulting eigenstates $\ket{\chi^\pm}$ ($\chi=\alp,\bet,\gam$, and $\del$) are shown in Table~\ref{eigen3} with
\begin{equation}
  C_\pm=\sqrt{\frac{1}{2}\left(1\pm\frac{J_1+J_2}{\sqrt{33J_1^2+2J_1J_2+J_2^2}}\right)}.
\end{equation}

\begin{table}[htbp]
  \centering\caption{Eigenstates $\ket{\chi^\pm}$ ($\chi=\alp,\bet,\gam$, and $\del$) of $\ham_c^\pri$}\label{eigen3}
  \begin{tabular}{l|cc} \hline
     & Configuration  \\ \hline
    $\ket{\alp^+}$ & $\ket{\uar\uar\uar}$　\\
    $\ket{\bet^+}$ &　$\frac{1}{\sqrt{2}}C_- \ket{\dar\uar\uar} -C_+ \ket{\uar\dar\uar} + \frac{1}{\sqrt{2}}C_-\ket{\uar\uar\dar}$ \\
    $\ket{\gam^+}$ & $-\frac{1}{\sqrt{2}}(\ket{\dar\uar\uar} - \ket{\uar\uar\dar})$ \\
    $\ket{\del^+}$ & $\frac{1}{\sqrt{2}}C_+\ket{\dar\uar\uar} +C_- \ket{\uar\dar\uar} +\frac{1}{\sqrt{2}}C_+ \ket{\uar\uar\dar}$ \\
    $\ket{\bet^-}$ &　 $\frac{1}{\sqrt{2}}C_-\ket{\uar\dar\dar} -C_+ \ket{\dar\uar\dar} + \frac{1}{\sqrt{2}}C_-\ket{\dar\dar\uar}$ \\
    $\ket{\gam^-}$ & $-\frac{1}{\sqrt{2}}(\ket{\uar\dar\dar} - \ket{\dar\dar\uar})$ \\
    $\ket{\del^-}$ & $\frac{1}{\sqrt{2}}C_+\ket{\uar\dar\dar} +C_- \ket{\dar\uar\dar} +\frac{1}{\sqrt{2}}C_+ \ket{\dar\dar\uar}$ \\
    $\ket{\alp^-}$ & $\ket{\dar\dar\dar}$ \\
    \hline
  \end{tabular}
\end{table}
The eigenenergy $\eps_{\chi^\pm}$ is given by
\begin{align}
  \eps_{\alp^\pm} &=  \frac{1}{8}(2J_1+J_2)\mp \frac{3}{2}H^\pri, \hspc \eps_{\gam^\pm} = -\frac{5}{8}J_2\mp \frac{H^\pri}{2}, \\
  \eps_{\bet^\pm} &= \frac{1}{8}\left(-J_1+2J_2-\sqrt{33J_1^2+2J_1J_2+J_2^2}\right)\mp \frac{H^\pri}{2}, \\
  \eps_{\del^\pm} &= \frac{1}{8}\left(-J_1+2J_2+\sqrt{33J_1^2+2J_1J_2+J_2^2}\right)\mp \frac{H^\pri}{2}.
\end{align}
$\ket{\chi^+}$ and $\ket{\chi^-}$ correspond to a Kramers doublet due to the time-reversal symmetry when $H^\pri=0$.
With applying magnetic field, every Kramers doublet splits off and degeneracy lifts [see \figref{3spin}(b)].
If interactions between the magnetic unit cells are completely classical and weaker than $\ham_c^\pri$, the ground state under the classical limit of inter-cluster interactions is given by the direct product of $\bet^+$ for $J_2/J_1=0.65$ as expected from \figref{3spin}(b). We note that the product state describes the $\Mn^\pri=1/3$ MP phase.
In the following, we use this ground state as an approximated $\Mn^\pri=1/3$ MP state, i.e. $\ket{1/3}\equiv \bigotimes_l \ket{\bet^+}_l$ where $l$ denotes the cluster number.

In order to obtain dispersion relations of $T^\pm$ excitations, we use a semi-classical approximation of inter-cluster interactions in subsequent subsections, where low-lying states in the cluster are taken into account in addition to the ground state.

\subsection{$T^+$ excitation}
\label{SecIVc}
To understand the $T^+$ excitation in the $\Mn^\pri=1/3$ MP phase of AFSC, we consider the fully spin-polarized $\ket{\alp^+}$ state as a low-lying state in the cluster, because $T^+$ increases magnetization from the ground state $\ket{1/3}$.
There are, of course, higher-order processes including other excited states, e.g., a mixed state of two $\ket{\alp^+}$ and one $\ket{\bet^-}$ in three magnetic unit cells, but such processes can be ignored because only two states, $\ket{\alp^+}$ and $\ket{\bet^+}$, contribute to low-energy excitations around the critical field $H_c^\pri=\left(3J_1-J_2+\sqrt{33J_1^2+2J_1J_2+J_2^2}\right)/8$ where $\ket{\alp^+}$ and $\ket{\bet^+}$ are degenerate [see \figref{3spin}(b)].
Therefore, we use the projection to these two states, $\qjc_+=\prod_l \br{\ket{\alp^+}_l\bra{\alp^+}_l+\ket{\bet^+}_l\bra{\bet^+}_l}$.
In the projected Hamiltonian, there is the constraint that $\ket{\alp^+}_l\bra{\alp^+}_l+\ket{\bet^+}_l\bra{\bet^+}_l=1$.
The operators $\tau_l^\dag=\ket{\alp^+}_l\bra{\bet^+}_l$, $\tau_l=\ket{\bet^+}_l\bra{\alp^+}_l$, and $n_l=\ket{\alp^+}_l\bra{\alp^+}_l$ correspond to a hard-core bosonic creation, annihilation, and number operators, respectively.
Thus, by using these operators with $\lam=1$, the projected Hamiltonian $\ham_+^\dpri=\qjc_+\ham^\pri\qjc_+$ except for the constant term is given by
\begin{equation}
  \ham_+^\dpri = t\sum_l (\tau^\dagger_l\tau_{l+1}+\mrm{H.c.})+ V\sum_l n_l n_{l+1} +\mu \sum_l n_l
\label{PM}
\end{equation}
with
\begin{align}
  t&=\frac{1}{4}C_-(J_1C_--2\sqrt{2}J_2C_+),\\
  V&=\frac{1}{8}\BR{J_1 C_-^4+J_2\br{C_-^2-C_+^2}^2},\\
  \mu&= \eps_{\bet^+}-\eps_{\alp^+}-\frac{1}{4}\br{J_1C_-^2+J_2C_+^2} .
\end{align}
If we define the Fourier transform of one-particle excited state as
\begin{equation}
  \ket{K} = \sum_l \ex^{\im Kl} \ket{\alp^+}_l \bigotimes_{l'(\neq l)}\ket{\bet^+}_{l'},
\end{equation}
this state is an eigenstate of the projected Hamiltonian $\ham_+^\dpri$ with dispersion energy
\begin{equation}
  \eps_K^\dpri = 2t \cos K + \mrm{const.} \label{dr}
\end{equation}
This eigenstate represents one-particle (hard-core boson) excited state obtained by the creation and annihilation operators, $\tau^\dag$ and $\tau$.

Analyzing the $T^+$ excitation in the $\Mn^\pri=1/3$ MP phase of AFSC, we use the relation $K=3q_x$ because the periodicity of AFSC is three times as long as that of the projected model (\ref{PM}).
The white line in Fig.~\ref{32}(c) exhibits the dispersion relation $\eps_K^\dpri$ including the constant energy in Eq.~\eqref{dr}.
The result reproduces well the dispersion relation of peak structure in the DSSF of AFSC.
This hard-core bosonic excitation is a collective mode of three-spin clusters (trimer), and thus we call it {\it trimeron} in this paper.
Furthermore, as explained in \secref{dssf2}, $T^+(q,\omega)$ in the $\Mn^\pri=1/3$ MP phase is qualitatively equivalent to $S^-(q_x,\omega)$ in the $\Mn=1/3$ MP phase and $S^+(q_x,\omega)$ in the $\Mn=2/3$ MP phase of the FSL.
We thus conclude that the origin of these DSSFs of FSL is the trimeron, that is, one-particle excitation of hard-core boson based on three-spin clusters.

\subsection{$T^-$ excitation}
\begin{figure*}[htbp]
 \begin{center}
  \includegraphics[height=55mm]{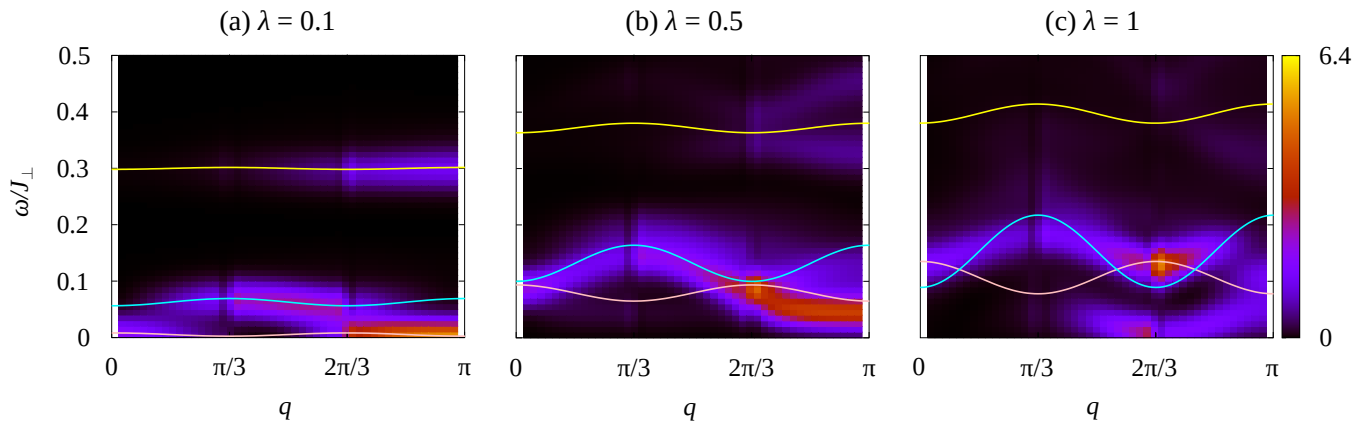}
 \end{center}
 \caption{
$T^-(q,\omega)$ in the effective model (\ref{Heff}) for the $\Mn^\pri=1/3$ MP phase of AFSC with various coupling ratios (a) $\lam=0.1$, (b) $\lam=0.5$, and (c) $\lam=1.0$.
 The dispersion relations colored by pink, blue, and yellow indicates $\bet^-$, $\gam^-$, and $\del^-$ modes, respectively,  in the projected Hamiltonian $\ham_-^\dpri$ without hybridization among these modes.
 }
 \label{evo}
\end{figure*}

The $T^-$ excitation in the $\Mn^\pri=1/3$ MP phase of AFSC is more difficult to understand than the $T^+$ excitation, because we have to take several states into account as low-lying states of the cluster.
However, our purpose in this section is to give an intuitive physical picture to explain the DSSF in \figref{16}(c).
Hence, quantitative reproduction is not necessary.
Here, we discuss the excitation through the CBSBS similar to the $T^+$ excitation.
We consider three excited states $\ket{\bet^-}$, $\ket{\gam^-}$, and $\ket{\del^-}$ as low-lying excited states of the cluster.
If interactions between clusters do not exist, the DSSF shows local excitations corresponding to $\ket{\bet^-}$, $\ket{\gam^-}$, and $\ket{\del^-}$, which induce three flat bands.
With adding the interactions as perturbation, the three bands become dispersive, indicating three modes of hard-core bosonic one-particle excitation (three trimerons) [see \figref{evo}(a)].
When the energy scale of interactions is larger than energy gaps between three excited states, theses modes are hybridized and split off.

Figure~\ref{evo} shows the $\lam$ dependence of $T^-(q,\ome)$ in the effective model (\ref{Heff}) for the $\Mn^\pri=1/3$ MP phase of AFSC.
Note that \figref{evo}(c) (the case of $\lambda = 1$) is the same as \figref{16}(c).
In Fig.~\ref{evo} we also plots dispersion relations of three modes, $\ket{\bet^-}$, $\ket{\gam^-}$, and $\ket{\del^-}$, which is obtained from the projected Hamiltonian $\ham_-^\dpri=\qjc_-\ham^\pri\qjc_-$ with $\qjc_-=\prod_l \br{\ket{\bet^+}_l\bra{\bet^+}_l+\sum_{\chi=\bet,\gam,\del}\ket{\chi^-}_l\bra{\chi^-}_l}$ but neglecting hybridization among the three modes for simplicity.
From \figref{evo}(c), we can easily imagine that, if we introduce hybridization effect, the $\bet^-$ and $\gam^-$ modes would be repulsively separated more around $q=2\pi/3$ and the $\gam^-$-originated mode would construct the lowest-energy excitations around $q=2\pi/3$ that are seen in $T^-(q,\ome)$. This speculation based on hybridized trimerons will explain the change of spectral distribution form $\lam=1$ to $\lam=0.5$, where the splitting of spectral weight around $q=2\pi/3$ disappears with decreasing $\lam$, because of the reduction of inter-cluster interactions controlling hybridization of  trimerons.

Based on these considerations in this subsection and \secref{SecIVc}, we conclude that $S^+(q_x,\omega)$ in the $\Mn=1/3$ MP phase and $S^-(q_x,\omega)$ in the $\Mn=2/3$ MP phase of the FSL originate from the hybridized trimerons.

\section{SUMMARY}\label{conclusion}
In this paper, we have studied magnetic excitations in the MP phases of the FSL, where the three types of antiferromagnetic interactions, $J_1$, $J_2$ and $J_\perp$, are taken into account as a leg nearest-neighbor, a leg second-neighbor, and a rung nearest-neighbor couplings of a two-leg ladder, respectively. This model exhibits three MPs at fractionalized finite magnetization $\Mn=M/M_\rsat=1/3$, $1/2$, and $2/3$ with respect to saturation magnetization $M_\rsat$. These MPs emerge robustly in the strong rung limit. Moreover, this condition allows us to map the model Hamiltonian into another quasi-spin model AFSC, by ignoring high-energy states in a rung. To obtain the intuitive physical picture of spin dynamics through the mapping to the AFSC, we have focused on the strong rung range.

We first have obtained magnetic excitations of the FSL by calculating DSSFs using DDMRG method. We have found that magnetic excitations in the MP phase are commensurate to the enlarged unit cell of the MP ground state. For the sake of comparison, we have also calculated DSSFs in the AFSC of quasi-spins, and have confirmed that the AFSC reproduces low-energy excitations of the FSL qualitatively. The $\Mn=1/3$, $1/2$, and $2/3$ MP states in the FSL correspond to the $\Mn^\pri=-1/3$, $0$, and $1/3$ MP states in the AFSC, respectively. The zero-magnetization ground state of the AFSC is well known as the dimerized state, so that elementary excitations are regarded as bound spinons. Therefore, we conclude that low-energy magnetic excitations of the FSL correspond to bound quasi-spinons based on the singlet and + triplet states of rung.

To clarify the $S^+$ ($S^-$) excitation in the $\Mn=2/3$ ($1/3$) MP state, we have additionally analyzed spin dynamics through the CBSBS in the AFSC. In the CBSBS, one cluster corresponds to an enlarged unit cell after spontaneously breaking of translational symmetry, and inter-cluster interactions are treated as perturbative effects as compared with the intra-cluster interaction. We have found a new quasi-particle mode as a hard-core bosonic excitation in the $\Mn=1/3$ MP state of the AFSC, which we call trimeron because it is a collective mode of spin trimers. This trimeron picture is common to the $S^+$ ($S^-$) excitation in the $\Mn=2/3$ ($1/3$) MP state of the FSL. On the other hand, the $S^-$ ($S^+$) excitation in the $\Mn=2/3$ ($1/3$) MP state is not well described as single trimeron mode. We have thus examined the intra-cluster interaction dependence of the DSSF. The obtained result indicates that inter-mode coupling enhanced by the inter-cluster interactions is crucial even for low-energy excitation. Actually, we have confirmed that two low-lying modes are hybridized in the excitation spectra with increasing the inter-cluster interactions, which are regarded as a hybridized trimerons. Consequently, we conclude that the $S^-$ ($S^+$) excitation in the $\Mn=2/3$ ($1/3$) MP state corresponds to the hybridized trimerons of quasi-spins.

Our results will be useful for understanding the low-energy physics in not only FSL materials such as \ce{BiCu2PO6}~\cite{abraham94,koteswararao07,mentre09,tsirlin10} and \ce{Li2Cu2O(SO4)2}~\cite{rousse17,sun15,vaccarelli17,vaccarelli19} but also weakly-coupled spin dimer compounds~\cite{kodama02,ruegg03,kimura16}, where magnetic excitations originating from the new quasi-particle can be clarified by inelastic neutron scattering experiments in a magnetic field. In such materials, spin dynamics we have clarified is also important for understanding the spin/heat transport~\cite{nagasawa14,jeon16,kawamata18} including their application to spintronics~\cite{uchida08,hirobe16}. Furthermore, the CBSBS is also useful for the analyses of low-energy excitations in MPs of various spin systems. We expect that new quasi-particles $N$-merons will be discovered as elementary excitations in the MPs, where the enlarged unit cell includes $N$ original cells after spontaneously breaking of translational symmetry.

\begin{acknowledgments}
We would like to thank H. Onishi, M. Mori, S. Maekawa, and K. Okamoto for fruitful discussions.
This work was partly supported by the creation of new functional devices and high-performance materials to support next-
generation industries (CDMSI) to be tackled by using a post-K computer, Grants-in-Aid for Young Scientists (B) (Grant No.16K17753), and the inter-university cooperative research program of IMR, Tohoku University.
Numerical computation in this work was carried out on the Supercomputer Center at Institute for Solid State Physics, University of Tokyo and the supercomputers at JAEA.
\end{acknowledgments}

\appendix
\section{MAGNETIZATION CURVE}
The DDMRG method usually requires numerous calculation resources as compared with the DMRG method for the ground state. As a consequence, the system size for DDMRG is smaller than that for DMRG.
In this paper, the system size of the FSL is set to be 48 rungs for DDMRG, in contrast to the system with 72 rungs~\cite{sugimoto15} and 144 rungs~\cite{sugimoto18} for DMRG.
Small system size prevents us from clear three MPs, but we have managed to find them with an appropriate parameter set by calculating magnetization curve in the 48-rung FSL with the DMRG method.

\begin{figure}[htbp]
 \begin{center}
  \includegraphics[width=80mm]{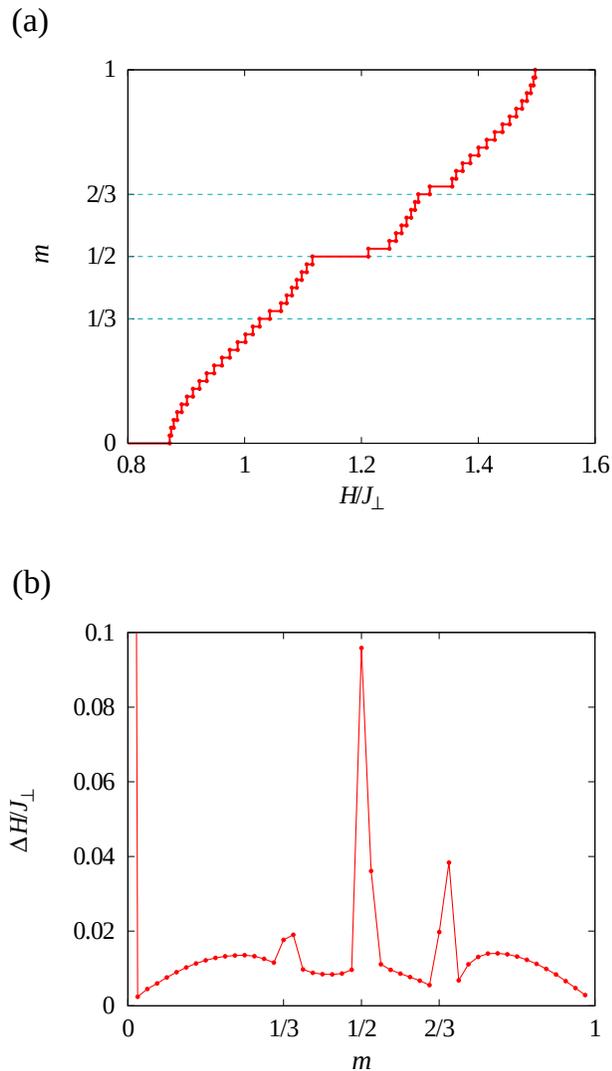}
 \end{center}
 \caption{(a) Magnetization curves normalized by saturation magnetization, $\Mn=M(H)/M_\rsat$, for the 48-rung FSL with $J_1/J_\perp=0.2,~J_2/J_1=0.65$.
 (b) Field width $\Delta H$ as a function of magnetization $\Mn$. The number of kept states in the DMRG is up to $500$ and the truncation error is less than $10^{-9}$.
 }
 \label{curve}
\end{figure}

Figure~\ref{curve}(a) shows magnetization curves with $J_1/J_\perp=0.2$ and $J_2/J_1=0.65$ in the 48-rung FSL.
Here, by using the Heaviside step function $\theta(H)$, magnetization curve is given by
\begin{align}
  M(H) = \sum_{M=1}^{M_\rsat-1} M\theta(H-H_{M-1,M}) \theta(H_{M,M+1}-H),
\end{align}
and the magnetic field where the ground states with two different magnetizations $M$ and $M+1$ are degenerate is obtained by $H_{M,M+1}\equiv E_{M+1} - E_{M}$ with the ground state energies $E_M$ and $E_{M+1}$ calculated by the DMRG method.
To confirm the presence of three MPs, we check the field width $\Delta H(M) = E_{M+1} - E_{M}$ at magnetization $M$ [see \figref{curve}(b)].
If the ground state with $M$ is gapless with respect to $S^+$ excitation, the field width converges to zero in the thermodynamical limit.
In \figref{curve}(b), we can see three jumps at $\Mn=1/3$, $1/2$, and $2/3$, indicating the presence of three MPs.
Therefore, we use this parameter set to calculate the DSSFs in this paper.

\section{MAGNETIC STRUCTURE IN $\Mn^\pri=\frac{1}{3}$ MP PHASE OF ASFC}
\begin{figure}[htbp]
 \begin{center}
  \includegraphics[width=75mm]{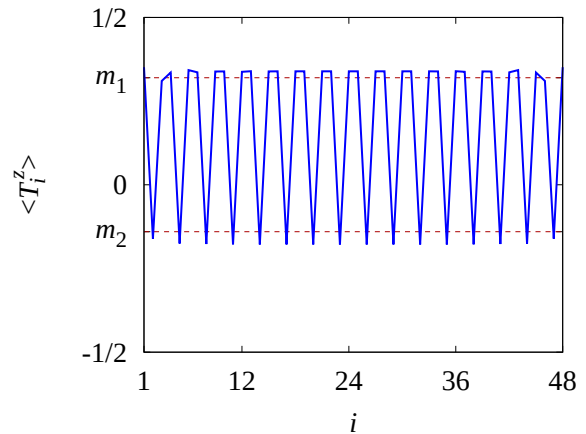}
 \end{center}
 \caption{Real-space expectation value of $z$ component of quasi-spin $T_i^z$ in the $\Mn^\pri=1/3$ MP phase.
 The red broken lines denote $m_1=\braket{\beta^+|T_1^z|\beta^+}=\braket{\beta^+|T_3^z|\beta^+}$ and $m_2=\braket{\beta^+|T_2^z|\beta^+}$. The $\beta^+$ state is the ground state in one-cluster (three quasi-spins) Hamiltonian (see Table~\ref{eigen3}).}
 \label{Sz16}
\end{figure}

In Sec.~IV, we use the CBSBS to explain the DSSFs in the $\Mn^\pri=1/3$ MP phase.
Although large energy gaps with $\Mn=1/3$ and $2/3$ are confirmed even in the 48-rung FSL, it is unclear whether spin configuration shows symmetry breaking with three times enlarged unit cell in the 48-site AFSC.
To confirm the enlarged magnetic unit cell in the ASFC, we calculate the expectation value of $z$ component of quasi-spin $T_i^z$ in the $\Mn^\pri=1/3$ MP phase as shown in \figref{Sz16}.
The period is found to be three times as long as the original unit cell.
Moreover, the expectation values are approximately equivalent to $m_1=\braket{\beta^+|T_1^z|\beta^+}=\braket{\beta^+|T_3^z|\beta^+}$ and $m_2=\braket{\beta^+|T_2^z|\beta^+}$, which are estimated in the ground state in one-cluster (three quasi-spins) Hamiltonian.
This result also supports the CBSBS approach.

\end{document}